\documentstyle[twocolumn,aps,epsf]{revtex}

\begin{document}
\title{Evaluating Capacities of Bosonic Gaussian Channels}
\author{A.~S. Holevo\thanks{Permanent address: Steklov Mathematical Institute,
Gubkina 8, 117966 Moscow, Russia} and R.~F. Werner }
\address{
Institut f\"ur Mathematische Physik, TU Braunschweig,
Mendelssohnstr.3, 38106 Braunschweig, Germany.
\\ Electronic Mail:
a.holevo@mi.ras.ru, r.werner@tu-bs.de}

\date{December 14, 1999}
\maketitle
\begin{abstract}
We show how to compute or at least to estimate various
capacity-related quantities for Bosonic Gaussian channels. Among
these are the coherent information, the entanglement assisted
classical capacity, the one-shot classical capacity, and a new
quantity involving the transpose operation, shown to be a general
upper bound on the quantum capacity, even allowing for finite
errors. All bounds are explicitly evaluated for the case of a
one-mode channel with attenuation/amplification and classical
noise.

\end{abstract}

\section{Introduction}
During the last years impressive progress was achieved in the
understanding of the classical and quantum capacities of quantum
communication channels (see, in particular, the papers \cite{Ben}
- \cite{Thap}, where the reader can also find further references).
It appears that a quantum channel is characterized by a whole
variety of different capacities depending both on the kind of the
information transmitted and the specific protocol used.

Most of this literature studies the properties of systems and
channels described in finite dimensional Hilbert spaces. Recently,
however, there has been a burst of interest (see e. g.
\cite{Braun}) in a new kind of systems, sometimes called
``continuous variable'' quantum systems, whose basic variables
satisfy Heisenberg's Canonical Commutation Relations (CCR). There
are two reasons for this new interest. On the one hand, such
systems play a central role in quantum optics, the canonical
variables being the quadratures of the field. Therefore some of
the current experimental realizations \cite{Kimble} of quantum
information processing are carried out in such systems. In
particular, the
 Bosonic Gaussian channels studied in this paper can be seen as
basic building blocks of quantum optical communication systems,
allowing to build up complex operations from ``easy, linear'' ones
and a few basic ``expensive, non-linear'' operations, such as
squeezers and parametric down converters.

The other reason for the interest in these systems is that in
spite of the infinite dimension of their underlying Hilbert spaces
they can be handled with techniques from finite dimensional linear
algebra, much in analogy to the finite dimensional quantum systems
on which the pioneering work on quantum information was done.
Roughly speaking this analogy replaces the density matrix by the
covariance matrix of a Gaussian state.  Then operations like the
diagonalization of density matrices, the Schmidt decomposition of
pure states on composite systems, the purification of mixed
states, the computation of entropies, the partial transpose
operation on states and channels, which are familiar from the
usual finite dimensional setup, can be expressed once again by
operations on finite dimensional matrices in the continuous
variable case. The basic framework for doing all this is not new,
and goes under heading ``phase space quantum mechanics'' or, in
the quantum field theory and statistical mechanics communities,
``quasi-free Bose systems'' \cite{demoen}. Both authors of this
paper have participated in the development of this subject a long
time ago \cite{Hol75,Hol82,wer}. In this paper, continuing
\cite{Sohma} and \cite{Hol98}, we make further contributions to
the study of information properties of linear Bosonic Gaussian
channels. We focus on the aspects essential for physical
computations and leave aside a number of analytical subtleties
related to infinite dimensionality and unboundedness unavoidably
arising in connection with Bosonic systems and Gaussian states.

The paper is organized as follows. In the Section~II we
recapitulate some notions of capacity, which are currently under
investigation in the literature, and what is known about them.
Naturally this cannot be a full review, but will be limited to
those quantities which we will evaluate  or estimate in the
subsequent sections. A new addition to the spectrum of
capacity-like quantities is discussed in Subsection~II.B: an upper
bound on the quantum capacity (even allowing finite errors), which
is both simple to evaluate and remarkably close to maximized
coherent information, a bound conjectured to be exact. In
Section~III we summarize the basic properties of Gaussian states.
Although our main topic is channels, we need this to get an
explicit handle on the purification operation, which is needed to
compute the entropy exchange, and hence all entropy based
capacities. Bosonic Gaussian channels are studied in Section~IV.
Here we introduce the techniques for determining the capacity
quantities introduced in Section~I, deriving general formulas
where possible. In the final Section~V we apply these techniques
to the case of a single mode channel comprising
attenuation/amplification and a classical noise. Some technical
points are treated in the Appendices.

\section{Notions of capacity}
\subsection{Basic entropy and information quantities} Consider a
general quantum system in a Hilbert space ${\cal H=H}_{Q}$. Its
states are given by density operators $\rho$ on ${\cal H}$. A {\it
channel\/} is a transformation $\rho \rightarrow T[\rho ]$ of
quantum states of the system, which is given by a completely
positive, trace preserving map on trace class operators. This view
of channels corresponds to the Schr{\"o}dinger picture. The
Heisenberg picture is given by the dual linear operator
$X\rightarrow T^*[X]$ on the observables $X$, which is
defined by the relation
\[
{\rm Tr}T[\rho ]X={\rm Tr}\rho T^*[X],
\]
and has to be completely positive and unit preserving (cf.
\cite{Hol72a}).

It can be shown (see e.g. \cite{Kraus}) that any channel in this
sense arises from a unitary interaction $U$ of the system with an
environment described by another Hilbert space ${\cal H}_{E}$
which is initially in some state $\rho_{E}$,
\[
T[\rho ]={\rm Tr}_{E}U\left( \rho \otimes \rho_{E}\right) U^*,
\]
where ${\rm Tr}_{E}$ denotes partial trace with respect to ${\cal H}_{E}$,
and vice versa. The representation is not unique, and the state $\rho_{E}$
can always be chosen pure, $\rho_{E}=|\psi_{E}\rangle \langle \psi_{E}|$.
The definition of the channel has obvious generalization to the case where
input and output are described by different Hilbert spaces.

Let us denote by
\begin{equation}\label{vNentropy}
 H(\rho )=-{\rm Tr}\rho \log \,\rho
\end{equation}
 the von Neumann entropy of a density operator $\rho $. We call $\rho$ the
input state, and $T[\rho ] $ the output state of the channel.
There are three important entropy quantities related to the pair
$(\rho ,T)$, namely, the entropy of the input state $H(\rho)$, the
entropy of the output state $H(T[\rho])$, and the entropy exchange
$H(\rho,T)$. While the definition and the meaning of the first two
entropies is clear, the third quantity is somewhat more
sophisticated. To define it, one introduces the {\it reference
system}, described by the Hilbert space ${\cal H}_{R}$, isomorphic
to the Hilbert space ${\cal H}_{Q}=$ ${\cal H}$ of the initial
system. Then according to \cite{Lin}, \cite{Sch}, there exists a
{\it purification} of the state $\rho $, i.e. a unit vector $|\psi
\rangle \in {\cal H}_{Q}\otimes {\cal H}_{R}$ such that
\[
\rho ={\rm Tr}_{R}|\psi \rangle \langle \psi |.
\]
The {\it entropy exchange } is then defined as
\begin{equation}\label{def-xchange}
  H(\rho ,T)=H\bigl((T\otimes {\rm id})[|\psi\rangle\langle\psi|]\bigr),
\end{equation}
that is, as the entropy of the output state of the dilated channel
$(T\otimes {\rm id})$ applied to the input which is purification of the state $\rho$.
Alternatively,
\[
H(\rho ,T)=H(\rho_{E}'),
\]
where $\rho_{E}'=T_{E}[\rho ]$ is the final state of the
environment, and the channel $T_E$ from ${\cal H}_{Q}$ to ${\cal
H}_{E}$ is defined as
\[
T_{E}[\rho ]={\rm Tr}_{Q}U\left( \rho \otimes \rho_{E}\right) U^*,
\]
provided the initial state $\rho_E$ of the environment is pure
\cite{Lin}, \cite{Sch}.

>From these three entropies one can construct several information
quantities. In analogy with classical information theory, one can
define {\it quantum mutual information }between the reference
system $R$ (which mirrors the input $Q$) and the output of the
system $Q'$ \cite{Lin}, \cite {Cer} as
\begin{eqnarray}\label{q-mutual}
   I(\rho ,T)&=&H(\rho_{R}')+H(\rho_{Q}')
                 -H(\rho_{RQ}')
\nonumber\\
             &=&H(\rho )+H(T[\rho ])-H(\rho ,T).
\end{eqnarray}
The quantity $I(\rho ,T)$ has a number of nice and ``natural''
properties, in particular, positivity, concavity with respect to
the input state $\rho $ and additivity for parallel channels
\cite{Cer}. Moreover, the maximum of $I(\rho, T )$ with respect to
$\rho$ was argued recently to be equal to the {\it
entanglement-assisted classical capacity} of the channel
\cite{Thap},\cite{Shor}, namely, the classical capacity of the
superdense coding protocol using the noisy channel $T$. It was
shown that this maximum is additive for parallel channels, the
one-shot expression thus giving the full (asymptotic) capacity.

It would be natural to compare this quantity with the (unassisted)
classical capacity $C(T)$ (the definition of which is outlined in
the next Subsection); however it is still not known whether this
capacity is additive for parallel channels. This makes us focus on
the one-shot expression, emerging from the coding theorem for
classical-quantum channels \cite{Hol}
\begin{equation}\label{oneshotunassist}
C_{1}(T)=\max \left[ H\Bigl( \sum_{i}p_{i}T\left[ \rho_{i}\right]
\Bigr)\right. \left.- \sum_{i}p_{i}H\left( T\left[
   \rho_{i}\right] \right)  \right] ,
\end{equation}
where the maximum is taken over all probability distributions
$\left\{ p_{i}\right\}$ and collections of density operators
$\left\{ \rho _{i}\right\}$ (possibly satisfying some additional
input constraints). $C_1(T)$ is equal to the capacity of $T$ for
classical information, if the coding is required to avoid
entanglement between successive inputs to the channel. The full
capacity is then attained as the length $n$ of the blocks, over
which encoding may be entangled goes to infinity, i.e.,
\begin{equation}\label{C-from-C1}
  C(T)=\lim_{n\to\infty}\ \frac1n C_1(T^{\otimes n}).
\end{equation}

An important component of $I(\rho ,T)$ is the {\it coherent
information}
\begin{equation}\label{def-coherent}
   J(\rho ,T)=H(T[\rho ])-H(\rho ,T),
\end{equation}
the maximum of which has been conjectured to be the (one-shot)
quantum capacity of the channel $T$ \cite{lloyd}, \cite{Sch}. Its
properties are not so nice. It can be negative, its convexity
properties with respect to $\rho $ are not known, and its maximum
was shown to be strictly superadditive for certain parallel
channels \cite{smo}, hence the conjectured full quantum capacity
may be greater than the one-shot expression, in contrast to the
case of the entanglement-assisted classical capacity. In this
paper we shall also compare this expression with a new upper bound
on the quantum capacity $Q(T)$ (as introduced e.g. in the next
Subsection).

\subsection{A general bound on quantum channel capacity}
In this Subsection we will establish a general estimate on the
quantum channel capacity, which will then be evaluated in the
Gaussian case, and will be compared with the estimates of coherent
information. Let us recall first a definition of the capacity
$Q(T)$ of a general channel $T$ for quantum information.
Intuitively, it is the number of qubits which can be faithfully
transmitted per use of the channel with the best possible error
correction. The standard of comparison is the ideal 1-qubit
channel ${\rm id}_2$, where ${\rm id}_n$ denotes the identity map
on the $n\times n$-matrices. Then the {\it quantum capacity}
$Q(T)$ of a channel $T$ (possibly between systems of different
type) is defined as the supremum of all numbers $c$, which are
``attainable rates'' in the following sense: {\it For any pair of
sequences $n_\alpha,m_\alpha$ with
$\lim_\alpha(n_\alpha/m_\alpha)= c$ we can find encoding
operations $E_\alpha$ and decoding operations $D_\alpha$ such
that}
\[
\Vert {\rm id}_2^{\otimes n_\alpha}- D_\alpha T^{\otimes
m_\alpha}E_\alpha\Vert_{{\rm cb}}\longrightarrow0.
\]
Here $\Vert \cdot\Vert_{{\rm cb}}$ is the so-called ``norm of
complete boundedness''\cite{Paulsen}, which is defined as the
supremum with respect to $n$ of the norms $\Vert(T\otimes {\rm
id}_n)\Vert$. It is equal to the ``diamond metric'' introduced in
\cite{AhaKitNi}. We use this norm because on the one hand, it
leads to the same capacity as analogous definitions based on other
error criteria (e.g., fidelities \cite{Sch,Barn}) and, on the
other hand, it has the best properties with respect to tensor
products, which are our main concern. In particular,
 $\Vert T\otimes S\Vert_{{\rm cb}}
 =\Vert T\Vert_{{\rm cb}}\cdot\Vert S\Vert_{{\rm cb}}$.
 Completely positive maps satisfy $\Vert T\Vert_{{\rm cb}}=\Vert F\Vert$,
where $F$ is the normalization operator determined by
${\rm Tr}(T[\rho])={\rm Tr}(\rho F)$. In particular, $\Vert T\Vert_{{\rm
cb}}=1$ for any channel. We also note another kind of capacity, in
which a much weaker requirement is made on the errors, namely
\begin{equation}  \label{errq}
\Vert {\rm id}_2^{\otimes n_\alpha}- D_\alpha T^{\otimes
m_\alpha}E_\alpha\Vert_{{\rm cb}}\leq\varepsilon<1
\end{equation}
for all sufficiently large $\alpha$, and some fixed $\varepsilon$.
We call the resulting capacity the {\it $\varepsilon$-quantum
capacity}, and denote it by $Q_{\varepsilon} (T)$. Of course,
$Q(T)\leq Q_{\varepsilon} (T)$, and by analogy with the classical
case (strong converse of Shannon's Coding Theorem) one would
conjecture that equality always holds.

The unassisted classical capacity $C(T)$ can be defined similarly
with the sole difference that both the domain of encodings $E$ and
the range of decodings $D$ should be restricted to the state space
of the Abelian subalgebra of operators diagonalizable in a fixed
orthonormal basis. In that case there is no need to use the
cb-norm, as it coincides with the usual norm. According to
recently proven strong converse to the quantum coding theorem
\cite{ogawa}, \cite{winter}, $C_{\varepsilon}(T)=C(T)$ where
$C_{\varepsilon}(T)$ is defined similarly to $Q_{\varepsilon}
(T)$.

The criterion we will formulate makes essential use of the transpose
operation, which we will denote by the same letter $\Theta $ in any system.
For matrix algebras, $\Theta $ can be taken as the usual transpose
operation. However, it makes no difference to our considerations, if any
other anti-unitarily implemented symmetry (e.g. time-reversal) is chosen. In
an abstract C*-algebra setting $\Theta $ is best taken as the ``op''
operation, which maps every algebra to its ``opposite''. This algebra has
the same underlying vector space, but all products $AB$ are replaced by
their opposite $BA$. Obviously, a commutative algebra is the same as
its opposite, so on classical systems $\Theta $ is the identity. Although
the transpose maps density operators to density operators, it is not an
admissible quantum channel, because positivity is lost, when coupling the
operation with the identity transformation on other systems, i.e., $\Theta $
is not {\it completely} positive. A similar phenomenon happens for the norm
of $\Theta $: we have $\Vert \Theta \Vert_{{\rm cb}}>1$ unless the system
is classical. In fact,
\begin{equation}\label{cbThetan}
\Vert \Theta_{n}\Vert_{{\rm cb}}=n,
\end{equation}
where $\Theta_{n}$ denotes the transposition on the $n\times n$-matrices
\cite{Paulsen}. We note that since we do not distinguish the transpose on
different systems in our notation, the observation that tensor products can
be transposed factor by factor is expressed by the equation $\Theta =\Theta
\otimes \Theta $. Moreover, although for a channel $T$, the operator
$T\Theta $ may fail to be completely positive, $\Theta T\Theta $ is again a
channel, and, in particular, satisfies $\Vert \Theta T\Theta \Vert_{{\rm cb}}=1$.

The main result of this Subsection is the estimate
\begin{equation}  \label{cbn-bound}
Q_\varepsilon(T)\leq \log\Vert T\Theta\Vert_{{\rm cb}} \equiv Q_\Theta(T),
\end{equation}
for any channel $T$. The proof is quite simple. Suppose
$n_\alpha/m_\alpha\rightarrow c\leq Q_\varepsilon(T)$, and encoding $E_\alpha$ and decoding
$D_\alpha$ are as in the definition of $Q_\varepsilon(T)$. Then by Equation~(
\ref{cbThetan}) we have
\begin{eqnarray}
2^{n_\alpha}
    &=& \Vert {\rm id}_2^{\otimes
                   n_\alpha}\Theta\Vert_{{\rm cb}} \ \leq
\nonumber \\
   &\leq&\Vert ({\rm id}_2^{\otimes n_\alpha}
       - D_\alpha T^{\otimes m_\alpha}E_\alpha)\Theta\Vert_{{\rm cb}}
\nonumber \\
    &&\qquad  +\Vert D_\alpha T^{\otimes
              m_\alpha}E_\alpha\Theta\Vert_{{\rm cb}}
\nonumber \\
   &\leq& \Vert \Theta_{2^{n_\alpha}}\Vert_{{\rm cb}}\
       \Vert {\rm id}_2^{\otimes n_\alpha}-
            D_\alpha T^{\otimes m_\alpha}E_\alpha\Vert_{{\rm cb}}
\nonumber \\
    &&\qquad  +\Vert D_\alpha (T\Theta)^{\otimes m_\alpha}\Theta
             E_\alpha\Theta\Vert_{{\rm cb}}  \nonumber \\
     &\leq& 2^{n_\alpha}\varepsilon
            + \Vert T\Theta\Vert_{{\rm cb}}^{m_\alpha},
\nonumber
\end{eqnarray}
where at the last inequality we have used that $D_\alpha$ and
 $\Theta E_\alpha\Theta$ are channels, and that the cb-norm is exactly tensor
multiplicative, so
 $\Vert X^{\otimes m}\Vert_{{\rm cb}}=\Vert X\Vert_{{\rm cb}}^m$.
Hence, by taking the logarithm and dividing by $m_\alpha$, we get
\[
\frac{n_\alpha}{m_\alpha}\log2 +\frac{\log(1-\varepsilon)}{m_\alpha}  \leq
\log\Vert T\Theta\Vert_{{\rm cb}}.
\]
If we take base $2$ logarithms, as is customary in information
theory, we have $\log2=1$. Then in the last inequality we can go
to the limit $\alpha\to\infty$, obtaining $c\leq Q_\Theta(T)$, and
Equation~(\ref{cbn-bound}) follows by taking the supremum over all
attainable rates $c$. Note that base $2$ logarithms are built into
the above definition of capacity, because we are using the ideal
qubit channel as the standard of comparison. This amounts only to
a change of units. If another base is chosen for logarithms is
chosen, this should also be done consistently in all entropy
expressions, and Equation~(\ref{cbn-bound}) holds once again
without additional constants.

The upper bound $Q_{\Theta }(T)$ computed in this way has some
remarkable properties, which make it a capacity-like quantity in
its own right. For example, it is exactly additive:
\begin{equation}
Q_{\Theta }(S\otimes T)=Q_{\Theta }(S)+Q_{\Theta }(T),  \label{capt-add}
\end{equation}
for any pair $S,T$ of channels, and satisfies the ``bottleneck inequality''
$Q_{\Theta }(ST)\leq \min \{Q_{\Theta }(S),Q_{\Theta }(T)\}$. Moreover, it
coincides with the quantum capacity on ideal channels:
$Q_{\Theta }({\rm id}_{n})=Q({\rm id}_{n})=\log
_{2}n$, and it vanishes whenever $T\Theta $ is completely
positive.
In particular, $Q_{\Theta}(T)=0$, whenever $T$ is {\it separable}
in the sense that it can be decomposed as $T=PM$ into a
measurement $M$ and a subsequent preparation $P$ based on the
measurement results. This follows immediately from the observation
that on classical systems transposition is the identity. Then
$P\Theta=\Theta P\Theta$ is a channel, and so is $MP\Theta$.
We note that $Q_{\Theta}$ is also closely related to the
entanglement quantity
$\log_{2}\Vert ({\rm id}\otimes \Theta)(\rho)\Vert_{1}$,
 i.e., the logarithm of the trace norm of the partial
transpose of the density operator, which enjoys analogous
properties.

\section{Quantum Gaussian states}
\subsection{Canonical Variables and Gaussian states}
In this Section we recapitulate some results from \cite{Hol75},
\cite{Sohma}, \cite{Hol98} for the convenience of the reader. Our
approach to quantum Gaussian states is based on the characteristic
function of the state which closely parallels classical
probability \cite{Hol82}, \cite{wer}, and is perhaps the simplest
and most transparent analytically. An alternative approach can be
based on the Wigner ``distribution function'' \cite{agar}.

Let $q_{j},p_{j}$ be the canonical observables satisfying the
Heisenberg CCR
\[
\lbrack q_{j},p_{k}]=i\delta_{jk}\hbar
I,\;\;[q_{j},q_{k}]=0,\;\;[p_{j},p_{k}]=0.
\]
We introduce the column vector of operators
\[
R=[q_{1},p_{1},\dots ,q_{s},p_{s}]^{T},
\]
the real column $2s$-vector $z=[x_{1},y_{1},\dots ,x_{s},y_{s}]^{T}$, and
the unitary operators in ${\cal H}$
\begin{eqnarray}\label{Weylop}
   V(z)&=&\exp \,i\sum_{j=1}^{s}(x_{j}q_{j}+y_{j}p_{j})\\
       &=&\exp \,i\,R^{T}z.\nonumber
\end{eqnarray}
These ``Weyl-operators'' satisfy the Weyl-Segal CCR
\begin{equation}
V(z)V(z')=\exp [\frac{i}{2}\Delta (z,z')]V(z+z'), \label{weyl}
\end{equation}
where
\begin{equation}\label{sympl-form}
 \Delta (z,z')
     =\hbar \sum_{j=1}^{s}(x_{j}'y_{j}-x_{j}y_{j}')
\end{equation}
is the canonical symplectic form. The space $Z$ of real $2s$-vectors
equipped with the form $\Delta (z,z')$ is what one calls a
{\it symplectic vector space}.
We denote by
\begin{equation}
\Delta =\left[
\begin{array}{ccccc}
0 & \hbar  &  &  &  \\
-\hbar  & 0 &  &  &  \\
&  & \ddots  &  &  \\
&  &  & 0 & \hbar  \\
&  &  & -\hbar  & 0
\end{array}
\right]   \label{delta}
\end{equation}
the $(2s)\times (2s)$-skew-symmetric {\it commutation matrix} of
components of the vector $R$, so that
\[
\Delta (z, z')= - z^{T}\Delta z'.
\]
 Most of the results
below are valid for the case where the commutation matrix is an
arbitrary (nondegenerate) skew-symmetric matrix, not necessarily
of the canonical form (\ref{delta}).

A density operator $\rho $ has {\it finite second moments }if
${\rm Tr}(\rho q_{j}^{2})<\infty$ and ${\rm Tr}(\rho p_{j}^{2})<\infty$
for all $j$. In this case one can define the vector {\it mean\/} and
the {\it correlation matrix} $\alpha$ by the formulas
\begin{equation}
 m={\rm Tr}\rho R\;;\;
 \alpha -\frac{i}{2}\Delta ={\rm Tr}(R-m)\rho (R-m)^{T}.
\label{alpha}
\end{equation}
The mean can be an arbitrary real vector. The correlation matrix $\alpha$
is real and symmetric. A given $\alpha$ is the correlation matrix
of some state if and only if it satisfies the {\it matrix uncertainty
relation}
\begin{equation}
\alpha -\frac{i}{2}\Delta \geq 0.  \label{n-s condition}
\end{equation}
We denote by $\Sigma \left( m,\alpha \right)$ the set of states
with fixed mean $m$\ and the correlation function $\alpha $.
The density operator $\rho $ is called {\it Gaussian}, if its {\it quantum
characteristic function} $\phi (z)={\rm Tr}\rho V(z)$ has the form
\begin{equation}\label{GaussianState}
  \phi (z)=\exp \left(i\,m^{T}z-\frac{1}{2}z^{T}\alpha z\right),
\end{equation}
where $m$ is a column ($2s$)-vector and $\alpha $\ is a real
symmetric $(2s)\times (2s)$-matrix. One then can show that $m$ is
indeed the mean, and $\alpha $ is the correlation matrix, and
(\ref{GaussianState}) defines the unique Gaussian state in
$\Sigma\left(m,\alpha\right)$. In what follows we will be
interested mainly in the case $m=0$.

The correlation matrix $\alpha$ describes a quadratic form rather
than an operator. Therefore its eigenvalues have no intrinsic
significance, and depend on the choice of basis in $Z$. On the
other hand, the operator $\widehat\alpha$ defined by $z^T\alpha
z=\Delta(z,\widehat\alpha z)$ has a basis free meaning. In matrix
notation it is $\widehat\alpha=\Delta^{-1}\alpha$. This operator
is always diagonalizable, and its eigenvalues come in pairs $\pm
i\gamma_j$. Diagonalizing this operator is essentially the same as
the {\it normal mode decomposition} of the phase space, when the
form $z^T\alpha z$ is considered as the Hamiltonian function of a
system of oscillators. It leads to a decomposition of the phase
space into two-dimensional subspaces, such that on the $j^{\rm
th}$ subspace we have (in some new canonical variables ${\tilde
q}_j, {\tilde p}_j$)
\begin{equation}\label{one-mode-alpha}
  \alpha=\hbar\left[
           \begin{array}{ll}\gamma_j & 0 \\0 & \gamma_j\end{array} \right]
\quad;\quad
  \Delta=\hbar\left[
           \begin{array}{ll}0&1 \\-1&0 \end{array} \right],
\end{equation}
and all terms between different blocks vanish. The matrix
uncertainty relation now requires $\gamma_j\geq1/2$, in which
equality holds iff $\rho_j$ is the pure (minimum-uncertainty)
state. Hence a general Gaussian state $\rho$ is pure if and only
if all $\gamma_j=1/2$, or
\begin{equation}
(\Delta ^{-1}\alpha )^{2}=-\ \frac{1}{4}I,  \label{pure}
\end{equation}
in which case $\Sigma \left( m,\alpha \right)$ reduces to a
single point.

\subsection{Gauge-invariant states}
We shall be interested in the particular subclass of Gaussian
states most familiar in quantum optics, namely, the states having
a P-representation
\begin{equation}
\rho =\int |\zeta \rangle \langle \zeta |\mu_{N}(d^{2s}\zeta )
\label{p-repr}
\end{equation}
where $\mu_{N}(d^{2s}\zeta )$ is the complex Gaussian probability
measure with zero mean and the correlation matrix $N$. (see e.g.
\cite{Hel}, Sec. V, 5. II). Here $\zeta \in {\bf C}^{s}$, $|\zeta
\rangle $ are the coherent vectors in ${\cal H}$, $a|\zeta \rangle
=\zeta |\zeta \rangle $, $N$ is positive Hermitian matrix such
that
\begin{equation}\label{def-N}
  N={\rm Tr}\left(a\,\rho \,a^{\dagger }\right)
\end{equation}
(we use here vector notations, where $a=[a_{1},\dots ,a_{s}]^{T}$
is a column vector and $a^{\dagger }=[a_{1}^{\dagger },\dots
,a_{s}^{\dagger }]$ is a row vector) and
$a_{j}=\frac{1}{\sqrt{2\hbar }}(q_{j}+ip_{j})$.

These states respect the natural complex structure in the sense
that they are invariant under the gauge transformations
$a\rightarrow a\exp (i\varphi)$. As shown in \cite{Sohma}, the
quantum correlation matrix of such states is
\[
\alpha =\hbar \left[
\begin{array}{ll}
{\rm Re}N+I/2 & -{\rm Im}N \\
{\rm Im}N & {\rm Re}N+I/2
\end{array}
\right] ,
\]
With Pauli matrices $I_{2},\sigma_{y}$, the real $2s\times 2s-$ matrices of
such form can be rewritten as complex $s\times s-$ matrices, by using the
correspondence
\[
\left[
\begin{array}{ll}
A & -B \\
B & A
\end{array}
\right] =I_{2}A+\sigma_{y}B\leftrightarrow A+iB,
\]
which is an algebraic isomorphism. Obviously,
\[
\frac{1}{2}{\rm Sp}\left[
\begin{array}{ll}
A & -B \\
B & A
\end{array}
\right] ={\rm Sp}(A+iB),
\]
where by ``${\rm Sp}$'' we denote the trace of matrices, as
opposed to the trace of Hilbert space operators, which is denoted
by ``Tr''. By using this correspondence, we have
\begin{equation}
\alpha \leftrightarrow \hbar (N+I/2),\qquad \Delta \leftrightarrow
-i\hbar I, \label{corr}
\end{equation}
and
\begin{equation}\label{arrow}
\Delta ^{-1}\alpha \leftrightarrow i(N+I/2).
\end{equation}

For the case of one degree of freedom we shall be interested in
the last Section, $N$ is just a nonnegative number and $\rho $ is
an {\it elementary} Gaussian state with the characteristic
function
\begin{equation}
\phi (z) =\exp \left[ -\frac{\hbar }{2}\left( N+\frac{1}{2}\right)
|z|^{2}\right] ,  \label{one-mode-cf}
\end{equation}
where we put $|z|^{2}=(x^{2}+y^{2})$. This state has correlation
matrix of the form (\ref{one-mode-alpha}) in the initial variables
$q,p$, with $\gamma = N+1/2$, and is just the temperature state of
the harmonic oscillator
\begin{equation}\label{onemode-rho}
   \rho_\gamma =\frac{1}{\gamma +1/2}\sum_{n =0}^{\infty }
    \left( \frac{\gamma -1/2}{\gamma +1/2}\right)^{n}\  |n\rangle\,\langle n|
\end{equation} in the number basis $|n\rangle$, with the mean photon number $N$.

\subsection{Computation of entropy}
To compute the von Neumann entropy of a general Gaussian state one
can use the normal mode decomposition. For a single mode, the
density operator $\rho_{j}$ with the correlation matrix
(\ref{one-mode-alpha}), setting $\gamma_j\equiv \gamma$ for
convenience, is unitarily equivalent to the state
(\ref{onemode-rho}). From this one readily gets the von Neumann
entropy $H(\rho_\gamma)$ by summation of the geometric series, and
for general Gaussian $\rho$ by summing over normal modes.

To write the result in compact form, one introduces the function
\begin{eqnarray}\label{def-g}
  g(x)&=&(x+1)\log (x+1)-x\log x,\quad x>0 \\
  g(0)&=&0.\nonumber
\end{eqnarray}
Then
\begin{equation}\label{H->g}
  H(\rho )= \sum_{j=1}^{s}\ g\left(|\gamma_{j}|-\frac{1}{2}\right),
\end{equation}
where $\gamma_j$ runs over all eigenvalue pairs $\pm i\gamma_j$ of
$\Delta^{-1}\alpha$.

One can also write this more compactly, using the following
notations, which we will also use in the sequel. For any
diagonalizable matrix $M=S{\rm diag}(m_{j})S^{-1}$, we put ${\rm
abs}(M)=S{\rm diag}(|m_{j}|)S^{-1}$, analogously for other
continuous functions on the complex plane.  Then
equation~(\ref{H->g}) can be written as \cite{Sohma}
\begin{equation}\label{abs}
 H(\rho )=\frac{1}{2}{\rm Sp}\ g\left( {\rm abs}(\Delta ^{-1}\alpha )
            -\frac{I}{2}\right) .
\end{equation}
For gauge-invariant state, by using (\ref{arrow}), this reduces to
the well-known formula
\[
H(\rho )={\rm Sp}\ g(N).
\]

\subsection{Schmidt Decomposition and Purification}
Forming a composite systems out of two systems described by
CCR-relations is very simple: one just joins the two sets of
canonical operators, making operators belonging to different
systems commute. The symplectic space of the composite system is a
direct sum $Z_{12}=Z_1\oplus Z_2$, which means that elements of
this space are pairs $(z_1,z_2)$ with components $z_i\in Z_i$. In
terms of Weyl operators one can write
$V_{12}(z_1,z_2)=V_{1}(z_1)\otimes V_{2}(z_2)$. By definition, the
symplectic matrix $\Delta_{12}$ is block diagonal with respect to
the decomposition $Z=Z_1\oplus Z_2$. However, the correlation
matrix $\alpha_{12}$ is block diagonal if and only if the state is
a product. The restriction of a bipartite Gaussian state $\rho$ to
the first factor is determined by the expectations of the Weyl
operators $V_1(z_1)\otimes{\bf1}=V_{12}(z_1,0)$, hence according to
(\ref{GaussianState}), by the the correlation matrix $\alpha_1$
with
 $z_1^T\alpha_1z_1=(z_1,0)^T\alpha_{12}(z_1,0)$, which is just the
first diagonal block in the block matrix decomposition
\begin{equation}\label{alpha1+2}
  \alpha_{12}=\left[
         \begin{array}{ll} \alpha_1 & \beta \\ \beta^T & \alpha_2
               \end{array}\right]\quad;\quad
   \Delta_{12}=\left[
          \begin{array}{ll} \Delta_1 & 0 \\ 0& \Delta_2
          \end{array}\right].
\end{equation}

As in the case of bipartite systems with finite dimensional
Hilbert spaces there is a canonical form for {\it pure} states of
the composite system, the Schmidt decomposition. Like the
diagonalization of a one-site density operator, it can be carried
out for Gaussian states at the level of correlation matrices. By
writing out equation~(\ref{pure}) in block matrix form, we find in
particular that
\begin{equation}\label{intertwine}
  (\Delta_1^{-1}\alpha_1)(\Delta_1^{-1}\beta)
  =(\Delta_1^{-1}\beta)(\Delta_2^{-1}\alpha_2).
\end{equation}
Thus $(\Delta_1^{-1}\beta)$ maps eigenvectors of
$(\Delta_2^{-1}\alpha_2)$ into eigenvectors of
$(\Delta_1^{-1}\alpha_1)$, with the same eigenvalue. Hence the
spectra of the restrictions are synchronized much in the same way
as in the finite dimensional case, and all the matrices
$\alpha_1,\alpha_2,\beta$ can be diagonalized simultaneously by a
suitable choice of canonical coordinates. Evaluating also the
diagonal part of Equation~(\ref{pure}), one gets an equation for
$\beta$, so that finally $\alpha_{12}$ is decomposed into blocks
corresponding to (a) pure components belonging to only one
subsystem, and not correlated with the other, and (b) blocks of a
standard form, which can be written like (\ref{alpha1+2}) with
$\alpha_1=\alpha_2=\alpha$, $\Delta_1=\Delta_2=\Delta$ from
(\ref{one-mode-alpha}), and
\begin{equation}\label{alpha12-pure}
   \beta=\hbar\sqrt{\gamma^2-\frac{1}{4}}\
            \left[\begin{array}{cc} 1&0\\0&-1
          \end{array}\right].
\end{equation}
The purification of a general Gaussian state can easily be read
off from this, by constructing such a standard form for every
normal mode. In order to write $\beta$ in operator form without
explicit reference to the normal mode decomposition, it is most
convenient to perform an appropriate reflection in the space
$Z_2$, by which $\beta$ becomes purely off-diagonal. Then we can
choose \cite{Hol72} $\Delta_1=\Delta=-\Delta_2$ and
$\alpha_2=\alpha_1=\alpha$, resulting in
\begin{equation}\label{beta12}
  \beta  = - \beta^T =\Delta\sqrt{-(\Delta ^{-1}\alpha )^{2}-I/4}.
\end{equation}
This also covers cases with $\beta=0$ for some modes, where, strictly
speaking no purification would have been necessary.
We thus have
\begin{equation}\label{da12}
    \Delta_{12}^{-1}\alpha_{12}=\left[\begin{array}{cc}
               \Delta ^{-1}\alpha & \sqrt{-(\Delta ^{-1}\alpha )^{2}-I/4}\\
               \sqrt{-(\Delta ^{-1}\alpha )^{2}-I/4} & - \Delta ^{-1}\alpha
               \end{array}\right].
\end{equation}
In the gauge-invariant case, we can use the correspondence
\begin{equation}
\Delta_{12}^{-1}\alpha_{12}\leftrightarrow \left[
\begin{array}{ll}
i(N+I/2) & \sqrt{N^{2}+N} \\ \sqrt{N^{2}+N} & -i(N+I/2)
\end{array}
\right],   \label{d-1a}
\end{equation} following from (\ref{corr}).

\section{Linear Bosonic Channels}
\subsection{Basic Properties}

The characteristic property of the channels considered in this
paper is their simple description in terms of phase space
structures. The key feature is that Weyl operators go into Weyl
operators, up to a factor. That is, the channel map in the
Heisenberg picture is of the form
\begin{equation}
   T^{*}(V'(z'))=V(K^T z') f(z'),  \label{linbos}
\end{equation}
where $K:Z\to Z'$ is a linear map between phase spaces with
symplectic forms $\Delta$ and $\Delta'$, respectively, and $f(z')$
is a scalar factor satisfying certain positive definiteness
condition to be discussed later. Because of the linearity of $K$,
such channels are called {\it linear Bosonic channels}
\cite{Hol72a}, and if, in addition, the factor $f$ is Gaussian,
$T$ will be called a {\it Gaussian channel}. In terms of
characteristic functions, Equation~(\ref{linbos}) can be written
as
\begin{equation}
\phi '(z')=  \phi (K^{T}z') f(z') ,  \label{ce}
\end{equation}
where  $\phi$ and $\phi'$ are the characteristic functions of
input state $\rho$ and output state $T[\rho]$, respectively.

We will make use of following key properties:

(A) The dual of a linear Bosonic channel transforms any polynomial
in the operators $R'$ into a polynomial in the $R$ of the same
order, provided the function $f$ has derivatives of sufficiently
high order. This property follows from the definition of moments
by differentiating the relation (\ref{linbos}) at the point $z'=0$.

(B) A Gaussian channel transforms Gaussian states into Gaussian
states. This follows from the definition of Gaussian state
and the relation (\ref{ce}).

(C) Linear Bosonic channels are covariant with respect to phase
space translations. That is if $\rho^z=V(-\Delta^{-1}z)\rho
V(-\Delta^{-1}z)^*$ is a shift of $\rho$ by $z$, $T[\rho]$ is
similarly shifted by $Kz$.

There is a dramatic difference in the capacities of a Gaussian
channel for classical as opposed to quantum information. Classical
information can be coded by using phase space translates of a
fixed state as signal states, so the output signals will also be
phase space translates of each other. Then no matter how much
noise the channel may add, if we take the spacing of the input
signals sufficiently large, the output states will also be
sufficiently widely spaced to be distinguishable with near
certainty. Therefore the unconstrained {\it classical capacity is
infinite}. The same would be true, of course, for a purely
classical channel with Gaussian noise. The classical capacity of
such channels becomes an interesting quantity, however, when the
``input power'' is taken to be constrained by a fixed value, which
we must take as one of the parameters defining the channel. Then
arbitrarily wide spacing of input signals is no longer an
alternative, because an intrinsic scale for this spacing has been
introduced.

The remarkable fact of quantum information on Gaussian channels is
that such an intrinsic scale is already there: it is given by
$\hbar$. As we will show, the quantum information capacity is
typically bounded even without an energy constraint. Loosely
speaking, although we send arbitrarily many well distinguishable
quantum signals through the channel, coherence in the form of
commutator relations is usually lost. Surprisingly, in spite of
the infinite classical capacity, the {\it capacity for quantum
information may be zero}, which means that even joining
arbitrarily many parallel channels with poor coherence properties
is not good enough for sending a single qubit. This phenomenon
will be explained in some detail in Section V.

The choice of the scalar function $f(z')$ is crucial for the
quantum transmission properties of the channel. Normalization of
$T$ requires that $f(0)=1$, and it is clear that $|f(z')|\leq1$
for all $z'$, from taking norms in (\ref{linbos}). Beyond that, it
is not so easy to see which choices of $f$ are compatible with the
complete positivity. If $f$ decays rapidly, $T^*$ maps most
operators to operators near the identity, which means that there
is very much noise. On the other hand, there will be a lower limit
to the noise, depending on the linear transformation $K$. Only
when $K$ is a symplectic linear map and $T$ is reversible, the
choice $f(z)\equiv1$ is possible. Otherwise, there is some
unavoidable noise.

There are two basic approaches to the determination of the
admissible functions $f$. The first is the familiar constructive
approach already used in Section II, based on coupling the system
to an environment, a unitary evolution and subsequent reduction to
a subsystem, with all of these operations in their linear
Bosonic/Gaussian form. Basically this reduces the problem to
linear transformations of systems of canonical operators. This
will be described in Subsection~B, and used for the calculation of
entropy exchange in Subsection~C. Alternatively, one can describe
the admissible functions $f$ by a twisted positive definiteness
condition, and this will be used for evaluating the bound
$C_\Theta(T)$ in Subsection~D.

\subsection{Bosonic channels via transforming canonical operators}

Let $R,R_{E}$ be vectors of canonical observables in
${\cal H},{\cal H}_{E}$, with the commutation matrices $\Delta,\Delta_{E}$.
Consider the linear transformation
\begin{equation}
R'=KR+K_{E}R_{E,}  \label{chan}
\end{equation}
where $K,K_{E}$ are real matrices (to simplicfy notations we write
$R, R_{E}$ instead of $R\otimes I_{E}, I\otimes R_{E}$ etc.) Then
the commutation matrix and the correlation with respect to $R'$
are computed via (\ref{alpha}) with $m=0$, namely
\[
\alpha'-\frac{i}{2}\Delta '={\rm Tr}R'\rho'
R^{\prime T}.
\]
We apply this to the special case $\rho'=\rho\otimes\rho_E$, where
$\rho_{E}$ and $\rho$ are density operators in
${\cal H}_{E}$ and ${\cal H}$ with the correlation
matrices $\alpha_{E}$ and $\alpha$, respectively.
Then using (\ref{chan}), we obtain
\begin{eqnarray}
\quad \Delta' &=&K\Delta K^{T}+K_{E}\Delta_{E}K_{E}^{T} \nonumber\\
\alpha' &=&K\alpha K^{T}+K_{E}\alpha_{E}K_{E}^{T}.  \label{trans}
\end{eqnarray}
Of course, the operators $R'$ need not form a complete set of
observables in ${\cal H}\otimes{\cal H}_{E}$, but in any case $\alpha'$
is the correlation matrix of a system containing just the
canonical variables $R'$, and it is this state which we will
consider as the output state of the channel.

For fixed state $\rho_E$ (state of the ``environment'') the
channel transformation taking the input state $\rho$ to the output
$\rho'$ is described most easily in terms of characteristic
functions:
\begin{equation}
\phi '(z')=  \phi (K^{T}z') \phi _{E}\left( K_{E}^{T}z'\right).  \label{cE}\\
\end{equation}
We can write this as a linear Bosonic channel in the form
(\ref{ce}) with
\begin{equation}\label{prefactor}
 f(z')=\phi _{E}\left( K_{E}^{T}z'\right)={\rm Tr}\rho_{E}V_{E}( K_{E}^{T}z')
\end{equation}
Thus the factor $f$ is expressed in terms of the characteristic
function of the initial state of the environment. Obviously, the
channel is Gaussian if and only if this state is Gaussian.

If we want to get the state of the environment after the channel
interaction, as required in the definition of exchange entropy, we have to
supplement the linear equation (\ref{chan}) by a similar equation
specifying the environment variables $R_E'$ after the interaction:
\begin{eqnarray*}
R' &=&KR+K_{E}R_{E,} \\
R_{E}' &=&LR+L_{E}R_{E,}
\end{eqnarray*}
Assuming that $Z=Z'$ and $\Delta'=\Delta$, one can always choose
$L,L_E$ such that the combined transformation is canonical, i.e.,
preserves the commutation matrix
\[
\left[
\begin{array}{cc}
\Delta  & 0 \\
0 & \Delta_{E}
\end{array}
\right] .
\]
Then the channel $T_{E}:\rho \rightarrow \rho_{E}'$ can be defined
by the relation
\[
T_{E}^{*}\left[ V_{E}(z_{E})\right] =V(L^{T}z_{E})\cdot \phi_{E}\left(
L_{E}^{T}z_{E}\right) ,
\]
and is thus also linear Bosonic.

\subsection{Maximization of mutual information}

The estimate for the entanglement assisted classical capacity
suggested by \cite{Shor} is the maximum of the quantum mutual
information (\ref{q-mutual}) over all states satisfying an
appropriate energy constraint. Evaluating this maximum becomes
possible by the following result:\footnote{ The proof of this
theorem was stimulated by a question posed to one of the authors
(A.H.) by P. W. Shor.} {\it Let $T$ be a Gaussian channel. The
maximum of the mutual information $I(\rho)$ over the set of states
$\Sigma \left( m,\alpha \right)$ with given first and second
moments is achieved on the Gaussian state.}

{\it Proof }(sketch). By purification (if necessary), we can
always assume that $\rho_{E}$ is pure Gaussian. Then we can write
\[
I(\rho )=H(\rho )+H(T[\rho ])-H(T_{E}[\rho ]).
\]
Let $\rho_{0}$ be the unique Gaussian state in $\Sigma \left( m,\alpha
\right) $. For simplicity we assume here that $\rho_{0}$ is nondegenerate.
The general case can be reduced to this by separating the pure component in
the tensor product decomposition of $\rho_{0}$. The function $I(\rho )$ is
concave and its directional derivative at the point $\rho_{0}$ is (cf. \cite
{Shor})
\begin{eqnarray}
    \nabla_{X}I(\rho_{0})
    &=&{\rm Tr}X(\ln \rho_{0}+I)+{\rm Tr}{\cal \ }T[X](\ln T[\rho_{0}]+I)
\nonumber\\
    &&\quad-{\rm Tr}T_{E}[X](\ln T_{E}[\rho_{0}]+I).
\nonumber
\end{eqnarray}
By using dual maps this can be modified to
\begin{eqnarray}
\nabla_{X}I(\rho_{0})
   &=&{\rm Tr}X\Bigl\{ \ln \rho_{0}+T^*[\ln T[\rho_{0}]]
\nonumber\\
   &&\qquad\quad    -T_{E}^*[\ln T_{E}[\rho_{0}]]+I\Bigr\} .
\label{h2}
\end{eqnarray}
Now by property (B) of Gaussian channels, the operators
$\rho_{0},T[ \rho_{0}],T_{E}[\rho_{0}]$ are (nondegenerate)
Gaussian density operators, hence their logarithms are quadratic
polynomials in the corresponding canonical variables (see Appendix
in \cite{Sohma}). By property (A) the expression in curly brackets
in (\ref{h2}) is again a quadratic polynomial in $R$, that is a
linear combination of the constraint operators in $\Sigma \left(
m,\alpha \right)$. Therefore, the sufficient condition (\ref{usl})
in the Appendix is fulfilled and $I(\rho)$ achieves its maximum at
the point $\rho_{0}\in \Sigma \left( m,\alpha \right) $.

This theorem implies that the maximum of $I(\rho)$ over a set of
density operators defined by arbitrary constraints on the first
and second moments is also achieved on a Gaussian density
operator. In particular, for an arbitrary quadratic Hamiltonian
$H$ the maximum of $I(\rho)$ over states with constrained mean
energy ${\rm Tr}\rho H$ is achieved on a Gaussian state. The
energy constraint is linear in terms of the correlation matrix:
\[
{\rm Sp}(\epsilon \alpha) \leq N,
\]
where $\epsilon $ is the diagonal energy matrix (see \cite{Sohma}).

When $\rho $ and $T$ are Gaussian, the quantities $H(\rho )$,
$H(T[\rho ]),H(\rho ,T)$ and  $I(\rho,T ), J(\rho,T )$ can in
principle be computed by using formulas (\ref{abs}),
(\ref{trans}), (\ref{da12}). Namely, $H(T[\rho ])$ is given by
formula (\ref{abs}) with $\alpha $ replaced by $\alpha '$ computed
via (\ref{trans}), and
\[
H(\rho,T)=\frac{1}{2}{\rm Sp}\,g\left({\rm abs}(\Delta_{12}^{-1}\alpha_{12}')
      -\frac{I}{2}\right) ,
\]
where
\begin{eqnarray}
\alpha_{12}'&=&\left[
\begin{array}{cc}
\alpha ' & K\beta \\
\beta^TK^T & \alpha
\end{array}
\right]\nonumber\\
\beta&=&\Delta \sqrt{-(\Delta ^{-1}\alpha )^{2}-I/4} \nonumber
\end{eqnarray}
is computed by inserting (\ref{chan}) into
\[
\alpha_{12}'-\frac{i}{2}\Delta_{12}'={\rm Tr}\left(
R',R_{2}\right) \rho \left( R',R_{2}\right) ^{T},
\]
where $R_{2}$ are the (unchanged) canonical observables of the
reference system.

 Alternatively, the entropy exchange can be
calculated as the output entropy $H(T_{E}[\rho])$ if an explicit
description of $T_{E}$ is available. We shall demonstrate this
method in the example of one-mode channels in the Appendix.

\subsection{Norms of Gaussian Transformations}
The transposition operation on a Bosonic system can be realized as
the time reversal operation, i.e., the operation reversing the
signs of all momentum operators, while leaving the position
operators unchanged. Obviously, the dual $T^{*}$ then takes Weyl
operators into Weyl operators. So transposition is just like a
linear Bosonic channel, albeit without the scalar factor $f(z') $
in Equation~(\ref{ce}). It is this factor which makes the
difference between positivity and complete positivity, and also
enters the norm $\Vert T\Vert_{{\rm cb}}$. In this Subsection we
will provide general criteria for deciding complete positivity and
computing the norm of general linear Bosonic transformations.

These are by definition the operators $T$ acting on Weyl operators
according to (\ref{linbos}) where  $f(z')$ is a scalar factor. We
will assume for simplicity (and in view of the applications in the
following sections) that the antisymmetric form
\begin{equation}
\Delta''(z_{1},z_{2})=\Delta' (z_{1},z_{2})-\Delta (K^T z_{1},K^T
z_{2}) \label{diffsymp}
\end{equation}
is non-degenerate. This makes the space $Z'$ with the form $\Delta
^{\prime \prime }$ into a phase space in its own right. With the
introduction of suitable canonical coordinates it becomes
isomorphic to $(Z,\Delta )$, so there exists an invertible linear
operator $A:Z\to Z$ such that $\Delta''(z_{1},z_{2})=\Delta
(A^{-1}z_{1},A^{-1}z_{2})$.

If $f$ is continuous and has sufficient decay properties (which
will be satisfied in our applications), there is a unique trace
class operator $\rho$ determined by the equation
\begin{equation}
{\rm Tr}(\rho V(z))=f(Az).  \label{rhof}
\end{equation}
{\it Then $T$ is completely positive if and only if $\rho $ is a
positive trace class operator}. This is a standard result in the
theory of quasi-free maps on CCR-algebras \cite{demoen}. It is
proved by showing that both properties are equivalent to a
``twisted positive definiteness condition'', namely the positive
definiteness of all matrices of the form
\[
M_{rs}=f(z_{r}-z_{s})\,\exp \bigl(-\frac{i}{2}\Delta'
(z_{r},z_{s})
        +\frac{i}{2}\Delta (K^T z_{r},K^T z_{s})\bigr),
\]
where $z_{1},\ldots ,z_{n}$ are an arbitrary choice of $n$ phase
space points.

If $\rho $ is a non-positive hermitian trace class operator, it
has a unique decomposition into positive and negative part: $\rho
=\rho _{+}-\rho_{-}$ such that $\rho_{\pm }\geq 0$, and
$\rho_{+}\rho_{-}=0$. Then $|\rho |=\rho_{+}+\rho_{-}$ and the
trace norm is $\Vert \rho \Vert _{1}={\rm Tr}(\rho_+)+{\rm
Tr}(\rho_-)$. Inserting $\rho_{\pm }$ into Equation~(\ref {rhof})
instead of $\rho $, we get two functions $f_{\pm }$ on phase space
and from Equation~(\ref{linbos}) two linear Bosonic
transformations $T_{\pm } $ with $T=T_{+}-T_{-}$. By the criterion
just proved, $T_{+}$ and $T_{-}$ are completely positive. Hence
\begin{eqnarray}
\Vert T\Vert_{{\rm cb}} &\leq &\Vert T_{+}\Vert_{{\rm cb}}+\Vert
T_{-}\Vert_{{\rm cb}}=\Vert T_{+}({\bf 1})\Vert +\Vert T_{-}({\bf
1})\Vert \nonumber  \\ &=&f_{+}(0)+f_{-}(0)={\rm Tr}(\rho_+)+{\rm
Tr}(\rho_-)
     =\Vert \rho \Vert_{1}\label{Tcb}
\end{eqnarray}

If the factor $f$ is a Gaussian, i.e.,
\begin{equation}
f(z)=\exp \bigl(-\frac{1}{2}\,z^{T}\beta z\bigr)  \label{Gaussf}
\end{equation}
for some positive definite matrix $\beta $, we can go one step
further. In this case we may decompose $\beta $ into normal modes
with respect to $\Delta ^{\prime \prime }$, which decomposes $T$
into a tensor product of one-mode Gaussian transformations
$T_{\ell }$, for each of which $\Vert T_{\ell }\Vert_{{\rm cb}}$
may be computed separately by the above method. This amounts to
computing the trace norm of the operator $\rho_{\gamma }$ given by
(\ref{onemode-rho}) with arbitrary positive $\gamma$. The absolute
value of $\rho _{\gamma }$ is obtained by taking absolute values
of all the eigenvalues, which still makes $\Vert \rho_{\gamma
}\Vert_{1}$ a geometric series:
\begin{equation}
\Vert \rho_{\gamma }\Vert_{1}=\frac{1}{\gamma +1/2}\sum_{n
=0}^{\infty }
    \left| \frac{\gamma -1/2}{\gamma +1/2}\right|^{n}
 =\max \{1,\frac{1}{2\gamma }\}.  \label{normrhogamma}
\end{equation}
This is all the information we need for the estimates of quantum
capacity in the following Section.

\section{The Case of One Mode}
\subsection{Attenuation/amplification channel with classical noise}

The channel we consider in this Section combines
attenuation/amplification \cite{Hol98} with additive classical
noise \cite{Shor}. It can also be described as the most general
one-mode gauge invariant channel, or in quantum optics
terminology, the most general one-mode channel not involving
squeezing. Channels of this type were also used in \cite{clalim}
as the basis for an analysis of the classical limit of quantum
mechanics.

 Let us consider the CCR with one degree of freedom
$a=\frac{1}{\sqrt{2\hbar} }(q+ip)$, and let $a_{0}$ be another
mode in the Hilbert space ${\cal H}_{0}={\cal H}_{E}$ of an
``environment''. Let the environment be initially in the vacuum
state, i.e., in the state with the characteristic function
(\ref{one-mode-cf}) with $N=0$. Let $\xi$ be a complex random
variable with zero mean and variance $N_{c}$ describing additive
classical noise in the channel. The linear attenuator with
coefficient $k<1$ and the noise $N_{c}$ is described by the
transformation
\[
  a'=ka+\sqrt{1-k^{2}}a_{0}+\xi
\]
in the Heisenberg picture. Similarly, the linear amplifier with
coefficient $k>1$ is described by the transformation
\[
a'=ka+\sqrt{k^{2}-1}a_{0}^{\dagger }+\xi .
\]
It follows that the corresponding transformations $T[\rho ]$ of
states in the Schr\"odinger picture both have the characteristic
function
\begin{eqnarray} \label{atten}
    {\rm Tr}&T[\rho ]&V(z)=  {\rm Tr}\rho V(kz)
\times\nonumber\\&&\quad\times
      \exp \left[-\frac{\hbar}{2}\bigl(
             |k^{2}-1|/2+N_{c}\bigr)\, |z|^{2}\right].
\end{eqnarray}

Let the input state $\rho$ of the system be the elementary
Gaussian with characteristic function (\ref{one-mode-cf}). Then the
entropy of $\rho$ is $H(\rho)=g(N)$. From
(\ref{atten}) we find that the output state $T[\rho]$ is again
elementary Gaussian with $N$ replaced by
\[
N'=k^{2} N + N'_0, \]
 where
\[
  N'_0=\max\{0,(k^2-1)\} +N_{c}
\]
is the value of the output mean photon number corresponding to the
input vacuum state. Then
\begin{equation}\label{out-1mode}
  H(T[\rho])=g(N').
\end{equation}

Now we calculate the exchange entropy $H(\rho,T)$. The (pure)
input state $\rho_{12}$ of the extended system ${\cal
H}_{1}\otimes {\cal H}_{2}$ is characterized by the $2\times
2-$matrix (\ref{d-1a}). The action of the extended channel
$(T\otimes {\rm id})$ transforms this matrix into
\[
\Delta_{12}^{-1}\widetilde{\alpha }_{12}\leftrightarrow \left[
\begin{array}{ll} i(N'+\frac{1}{2}) & k\sqrt{N(N+1)} \\
k\sqrt{N(N+1)} & -i(N+\frac{1}{2}) \end{array} \right]. \]
From
formula (\ref{H->g}) we deduce
$H(\rho,T)=g(|\lambda_{1}|-\frac{1}{2})+g(|\lambda_{2}|-\frac{1}{2})$,
where $\lambda_{1},\lambda_{2}$ are the eigenvalues of the complex
matrix in the right-hand side. Solving the characteristic equation
we obtain \begin{equation} \lambda_{1,2}=\frac{i}{2}\left(
(N'-N)\pm D\right) , \label{eig} \end{equation} where
$D=\sqrt{\left( N+N'+1\right) ^{2}-4k^{2}N(N+1)}$. Hence
\begin{eqnarray}\label{exch-1mode}
  H(\rho,&&T)=\\
  &&=g\left( \frac{D+N'-N-1}{2}\right)
    +g\left(\frac{D-N'+N-1}{2 }\right).\nonumber
\end{eqnarray}

Now using the theorem of Section 5, we can calculate  the quantity
\[
C_{e}(T)= I(\rho ,T)=H(\rho ) + H(T[\rho]) - H(\rho ,T)
\]
 as a function of the
parameters $N,k,N_{c}$, and try to compare it
with the one-shot unassisted classical capacity of the
channel $C_1 (T)$ given by expression (\ref{oneshotunassist})
where the maximum is taken over all probability distributions
$\left\{ p_{i}\right\} $ and the collections of density operators
$\left\{ \rho _{i}\right\} $, satisfying the power constraint
$\sum_{i}p_{i}$Tr$\rho _{i}a^{\dagger }a\leq N$. It is quite
plausible, but not yet proven that this maximum is achieved on
coherent states with the Gaussian probability density $p(z)=\left(
\pi N\right) ^{-1}\exp \left( -|z|^{2}/N\right) $, giving the
value
\[
\underline{C}_{1}(T)=g\left( N'\right) -g\left( N_{0}'\right).
\]

The ratio
\begin{equation}\label{gain}
   G=\frac{C_{e}}{\underline{C}_{1}}
\end{equation}
then gives at least an upper
bound for the {\it gain} of using entanglement-assisted versus
unassisted classical capacity. In particular, when the signal mean
photon number $N$ tends to zero while $N'_0 >0$,
\begin{eqnarray*}
\underline{C}_{1}(T)&\sim& N k^2 \log\left(\frac{N'_0
                     +1}{N'_0}\right),\\
              C_e (T)&\sim& - N \log N /(N'_0 +1),
\end{eqnarray*}
and $G$ tends to infinity as $ - \log N$.

The plots of $G$ as function of $k$ for $N_{c}=0$, and as a
function of $N_c$ for $k=1$ are given in Figure~1 and Figure~2,
respecitively. The behavior of the entropies
$H(T[\rho]),H(\rho,T)$ as functions of $k$ for $N_{c}=0$ is clear
from Figure~3. For all $N$ the coherent information
$H(T[\rho])-H(\rho,T)$ turns out to be positive for $k>1/\sqrt{2}$
and negative otherwise. It tends to $-H(\rho )$ for $k\rightarrow
0$, is equal to $H(\rho)$ for $k=1$, and quickly tends to zero as
$k\rightarrow \infty$ (see Figure~4).

\subsection{Estimating the quantum capacity}
Going back to the upper bound for quantum capacity in Section IV,
we see that  $T$ is given by equation~(\ref {linbos}) with $Kz=kz$
and
\[
f(z)=\exp (-\frac{(|k^{2}-1|/2+N_{c})}{2}\ |z|^{2}).
\]
Then $\Delta ^{\prime \prime }=(1-k^{2})\Delta $, and the operator
$A$ mapping the symplectic form $\Delta ^{\prime \prime }$ to the
standard form $\Delta $ is multiplication by $\sqrt{|k^{2}-1|}$,
combined for $k>1$ with a mirror reflection to change the sign.
This leaves
\begin{equation}
f(Az)=\exp \left(-\frac{(|k^{2}-1|/2+N_{c})}{2|k^{2}-1|}\ |z|^{2}\right),
\label{fLT}
\end{equation}
i.e., $\rho =\rho_{\gamma }$ with equations~(\ref{onemode-rho})
and (\ref{rhof}), where $\gamma =1/2+N_{c}/|k^{2}-1|$. This is the
verification of the complete positivity of $T$ by the methods of
the above section. Of course, this is strictly speaking
unnecessary, because $T$ was constructed explicitly as a
completely positive operator in terms of its dilation in
Subection~IV.A.

But let us now consider $T\Theta $. It is also a Bosonic linear
transformation, in which $\Theta $ only has the effect of changing
the sign of the symplectic form, without changing $f$. Thus
$\Delta ^{\prime \prime }=(1+k^{2})\Delta $, and
\[
f(Az)=\exp \left( -\frac{(|k^{2}-1|/2+N_{c})}{2|k^{2}+1|} |z|^{2}
\right) .
\]
which seems like a rather minor change over Equation~(\ref{fLT}). However,
we now get $\rho =\rho_{\gamma }$ with $\gamma
=(|k^{2}-1|/2+N_{c})/(k^{2}+1)$ which is not necessarily $\geq 1/2$, so
$T\Theta $ is not necessarily completely positive. Taking the logarithm of
Equation~(\ref{normrhogamma}) we get
\begin{eqnarray}
Q_{\Theta }(T)&\leq& \max\{0,
\nonumber\\
    &&\ \log_{2}(k^{2}+1)-\log_{2}(|k^{2}-1|+2N_{c})\}.
\label{1modBound}
\end{eqnarray}
In particular, for $\gamma \geq 1/2$, i.e., for $N_{c}\geq
(|k^{2}+1|-|k^{2}-1|)/2 =\max \{1,k^{2}\}$, the capacities
$Q_{\Theta }(T)$, and hence $Q_{\varepsilon }(T)$ and $Q(T)$ all
vanish.

This upper bound on quantum capacity is interesting to compare
with the quantity $Q_{G}(T)=\sup J(\rho ,T)$, where $J(\rho
,T)=H(T[\rho])-H(\rho ,T)$, and the supremum is taken over all
{\it Gaussian} input states. Since the coherent information
\begin{eqnarray}
  J(\rho ,T)&=& g(N')-g\left( \frac{D+N'-N-1}{2}\right)-
  \nonumber\\
    &&\quad -g\left( \frac{D-N'+N-1}{2}\right)
  \label{coh-1mode}
\end{eqnarray}
increases  with the input
power $N$, we obtain
\begin{eqnarray}\label{QG}
  Q_{G}(T)&=&\lim_{N\rightarrow \infty }J(\rho ,T)\\
          &=&\log k^{2}-\log |k^{2}-1|-
               g\left( N_{c}/|k^{2}-1|\right),\nonumber
\end{eqnarray}
which is in a good agreement with the upper bound
(\ref{1modBound})(see Figure 4).

\acknowledgments{A.H. appreciates illuminating discussion of
fragments of the unpublished paper \cite{Shor} with C. H. Bennett
and P. W. Shor. He acknowledges the hospitality of R. W. in the
Institute for Mathematical Physics, Technical University of
Braunschweig, which he was visiting with an A. von Humboldt
Research Award.}

\appendix
\section{Minimizing convex function of a density operator.}

There is a useful lemma in classical information theory which
gives necessary and sufficient conditions for the global minimum
of a convex function of probability distributions in terms of the
first partial derivatives. The lemma is based on general
Kuhn-Tucker conditions and can be generalized to functions
depending on density operators rather than probability
distributions.

Let $F$ be a convex function on the set of density operators
$\Sigma $, and $\rho_{0}$ a density operator. In order $F$ to
achieve minimum on $\rho _{0}, $ it is necessary and sufficient
that for arbitrary density operator $\sigma $ the convex function
$F((1-t)\rho_{0}+t\sigma )$ of the real variable $t$ achieves
minimum at $t=0$. For this, it is necessary and sufficient that
\begin{equation}
\nabla_{X}F(\rho_{0})\equiv \left. \frac{d}{dt}\right|_{t=0}F((1-t)\rho
_{0}+t\sigma )\geq 0,  \label{nsc}
\end{equation}
where $X=\sigma -\rho_{0}$, and $\nabla_{X}F(\rho_{0})$ is the
directional derivative of $F$ in the direction $X$, assuming that
the derivatives exist. If $\sigma =\sum_{i}p_{i}$ $\sigma_{i}$,
then $\nabla _{X}F(\rho_{0})=\sum_{i}p_{i}$
$\nabla_{X_{i}}F(\rho_{0})$, where $X_{i}=\sigma_{i}-$ $\rho_{0}$.
Therefore it is necessary and sufficient that (\ref{nsc}) holds
for pure $\sigma $.

If $(1-t)\rho_{0}+t\sigma \geq 0$ for small negative $t$, then we
say that the direction $\overrightarrow{\sigma \rho_{0}}$ is {\it
inner}. In that case (\ref{nsc}) takes the form
\begin{equation}
\nabla_{X}F(\rho_{0})=0.  \label{nsp}
\end{equation}
If $\rho_{0}$ is nondegenerate, then the direction
$\overrightarrow{\sigma \rho_{0}}$ is inner for arbitrary pure
$\sigma $ in the range of $\sqrt{\rho_{0}}$, and the necessary and
sufficient condition for the minimum is that (\ref{nsp}) holds for
arbitrary such $\sigma $.

Let $A_{i},i=1,\dots ,r$ be a collection of selfadjoint {\it
constraint operators}. Assume that for some real constants
$\lambda_{i}$
\begin{equation}
\nabla_{X}F(\rho_{0})={\rm Tr}X\sum_{i}\lambda_{i}A_{i}.  \label{usl}
\end{equation}
It follows that the convex function $F(\rho )-{\rm Tr}\rho \sum_{i}\lambda
_{i}A_{i}$ achieves minimum at the point $\rho_{0}$, hence the function
$F(\rho )$ achieves minimum at the point $\rho_{0}$ under the constraints
${\rm Tr\ }\rho A_{i}={\rm Tr\ }\rho_{0}A_{i},\quad
i=1,\dots,r$ .

\section{Quantum signal plus classical noise.}

Let us consider CCR with one degree of freedom described by one
mode annihilation operator $a=\frac{1}{\sqrt{2\hbar} }(q+ip)$, and
consider the transformation
\[
a'=a+\xi ,
\]
where $\xi $ is a complex random variable with zero mean and
variance $N_{c}$. This is a transformation of the type
(\ref{chan}) with $\Delta_{E}=0$, which describes quantum mode in
classical Gaussian environment. The action of the dual channel is
\[
T^*[f(a,a^{\dagger })]
   =\int f(a+z,(a+z)^{\dagger })\mu_{N_{c}}(d^{2}z),
\]
where $z=\frac{1}{\sqrt{2\hbar} }(x+iy)$ is now complex variable,
and $\mu_{N_{c}}(d^{2}z)$ is complex Gaussian probability measure
with zero mean and variance $N_{c}$, while the channel itself can
be described by the formula
\begin{equation}
T[\rho ]=\int D(z)\rho D(z)^*\mu_{N_{c}}(d^{2}z),  \label{spn}
\end{equation}
where $D(z)=\exp i\left( za^{\dagger }-\bar{z}a\right) $ is the
displacement operator.

The entanglement-assisted classical capacity of the channel
(\ref{spn}) was first studied in \cite{Shor} by using rather
special way of purification and the computation of the entropy
exchange. A general approach following the method of \cite{Hol98}
was described in Sections IV-V; here we give an alternative
solution based on the computation of the environment entropy.

For this we need to extend the environment to a quantum system in
a pure state. Consider the environment Hilbert space ${\cal
H}_{E}=L^{2}(\mu_{N_{c}})$ with the vector $|\Psi_{0}\rangle$
given by the function identically equal to 1. The tensor product
${\cal H}\otimes {\cal H}_{E}$ can be realized as the space
$L_{{\cal H}}^{2}(\mu_{N_{c}})$ of $\mu_{N_{c}}$-square integrable
functions $\psi (z)$ with values in ${\cal H}$. Define the unitary
operator $U$ in ${\cal H}\otimes {\cal H}_{E}$ by
\[
(U\psi )(z)=D(z)\psi (z).
\]
Then
\[
T[\rho]={\rm Tr}_{{\cal H}_{E}}U\left( \rho \otimes
|\Psi_{0}\rangle\langle\Psi _{0}|\right) U^*,
\]
while
\[
T_{E}[\rho ]={\rm Tr}_{{\cal H}}U\left(
   \rho \otimes |\Psi_{0}\rangle\langle\Psi_{0}|\right) U^*.
\]
This means that $T_{E}[\rho ]$ is an integral operator in
$L^{2}(\mu _{N_{c}})$ with the kernel
\begin{eqnarray*}
  K(z,z')&=&{\rm Tr}D(z)\rho_{0}D(z')^*\\
    &=&\exp (i\Im \bar{z}'z-(E+1/2)|z-z'|^{2}).
\end{eqnarray*}
Let us define unitary operators $V(z_{1},z_{2})$ in $L^{2}(\mu_{N_{c}})$
by
\begin{eqnarray*}
   V(z_{1},z_{2})&\psi(z)&= \psi (z+z_{2})\times\\
       &&\ \times\exp
   \left[ i\Re \overline{z_{1}}(z+\frac{z_{2}}{2})-\frac{1}{N_{c}}
   \Re \overline{z_{2}}(z+\frac{z_{2}}{2})\right] .
\end{eqnarray*}
The operators $V(z_{1},z_{2})$ satisfy Weyl-Segal CCR with two degrees of
freedom with respect to the symplectic form
\[
\Delta ((z_{1},z_{2}),(z_{1}',z_{2}'))
   =\Re \left( \bar{z}_{1}'z_{2}-\bar{z}_{1}z_{2}'\right) .
\]
Passing over to the real variables $x,y$ one finds the
corresponding commutation matrix
\[
\Delta_{E}=\hbar\left[
\begin{array}{cccc}
0 & 0 & -1 & 0 \\
0 & 0 & 0 & -1 \\
1 & 0 & 0 & 0 \\
0 & 1 & 0 & 0
\end{array}
\right] .
\]

The characteristic function of the operator $T_{E}[\rho_{0}]$ is
\[
{\rm Tr}T_{E}[\rho_{0}]V(z_{1},z_{2})
    =\int \left. V(z_{1},z_{2})K(\check{z},z)\right|_{\check{z}=z}
       \mu_{N_{c}}(d^{2}z),
\]
where $V(z_{1},z_{2})$ acts on $K$ as a function of the argument $\check{z}$.
Evaluating the Gaussian integral, we obtain that it is equal to
\[
\exp \left[ -\frac{1}{4}\left( N_{c}|z_{1}|^{2}+2N_{c}
\Im \bar{z}_{1}z_{2}+\frac{D^{2}}{N_{c}}|z_{2}|^{2}\right) \right] ,
\]
(where now $D=\sqrt{(N_c + 1)^2 + 4N_c N}$), which is Gaussian
characteristic function with the correlation matrix
\[
\alpha_{E}'=\frac{\hbar}{2}\left[
\begin{array}{cccc}
N_{c} & 0 & 0 & N_{c} \\
0 & N_{c} & -N_{c} & 0 \\
0 & -N_{c} & \frac{D^{2}}{N_{c}} & 0 \\
N_{c} & 0 & 0 & \frac{D^{2}}{N_{c}}
\end{array}
\right] .
\]
Thus
\[
\ \Delta_{E}^{-1}\alpha_{E}'=\frac{1}{2}\left[
\begin{array}{cccc}
0 & -N_{c} & \frac{D^{2}}{N_{c}} & 0 \\
N_{c} & 0 & 0 & \frac{D^{2}}{N_{c}} \\
-N_{c} & 0 & 0 & -N_{c} \\
0 & -N_{c} & N_{c} & 0
\end{array}
\right] .
\]
By using Pauli matrix $\sigma_{y}$, we can write it as
\begin{eqnarray*}
\frac{1}{2}\left[
\begin{array}{cc}
-i\sigma_{y}N_{c} & \frac{D^{2}}{N_{c}} \\
-N_{c} & -i\sigma_{y}N_{c}
\end{array}
\right] &&=\\
=\frac{1}{2}\left[
\begin{array}{cc}
I & 0 \\
0 & \sigma_{y}
\end{array}
\right]&& \left[
\begin{array}{cc}
-i\sigma_{y}N_{c} & \sigma_{y}\frac{D^{2}}{N_{c}} \\
-\sigma_{y}N_{c} & -i\sigma_{y}N_{c}
\end{array}
\right] \left[
\begin{array}{cc}
I & 0 \\
0 & \sigma_{y}
\end{array}
\right] ,
\end{eqnarray*}
hence the absolute values of the eigenvalues of
$\Delta_{E}^{-1}\alpha_{E}'$ are the same as that of the matrix
\[
\left[
\begin{array}{cc}
iN_{c} & -\frac{D^{2}}{N_{c}} \\
N_{c} & iN_{c}
\end{array}
\right] ,
\]
which coincide with (\ref{eig}) in the case $k=1$.

\section*{Figures}

\begin{figure}\label{Fig-Gofk}

\epsfxsize=8cm \epsfbox{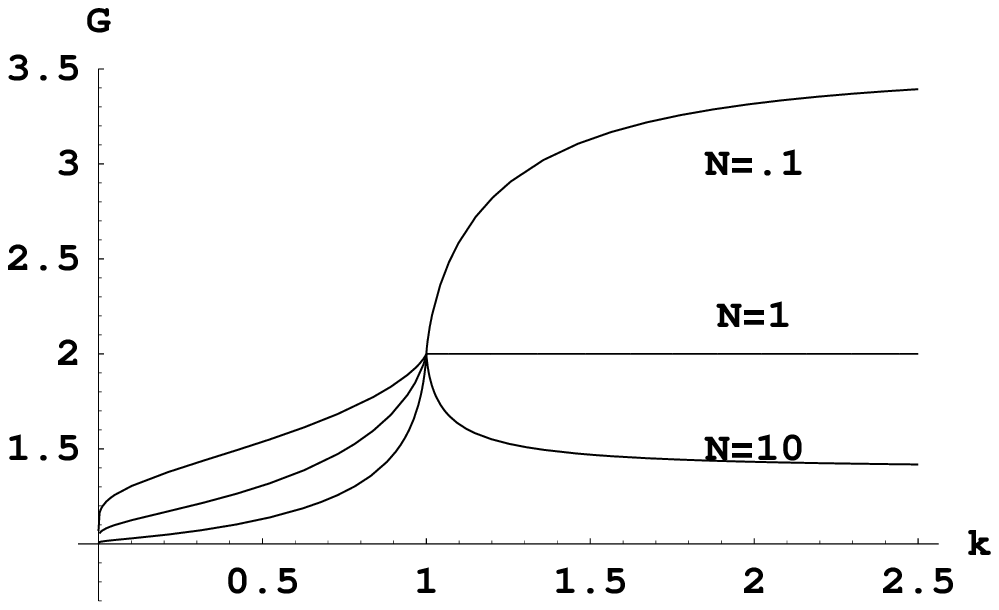} \caption{{\it Gain of
entanglement assistance.}}\small\narrower\noindent Gain
(\ref{gain}) as a function of $k$ with $N_c=0$. Parameter=input
noise $N$.
\end{figure}

\begin{figure}\label{Fig-GofNc}
\epsfxsize=8cm \epsfbox{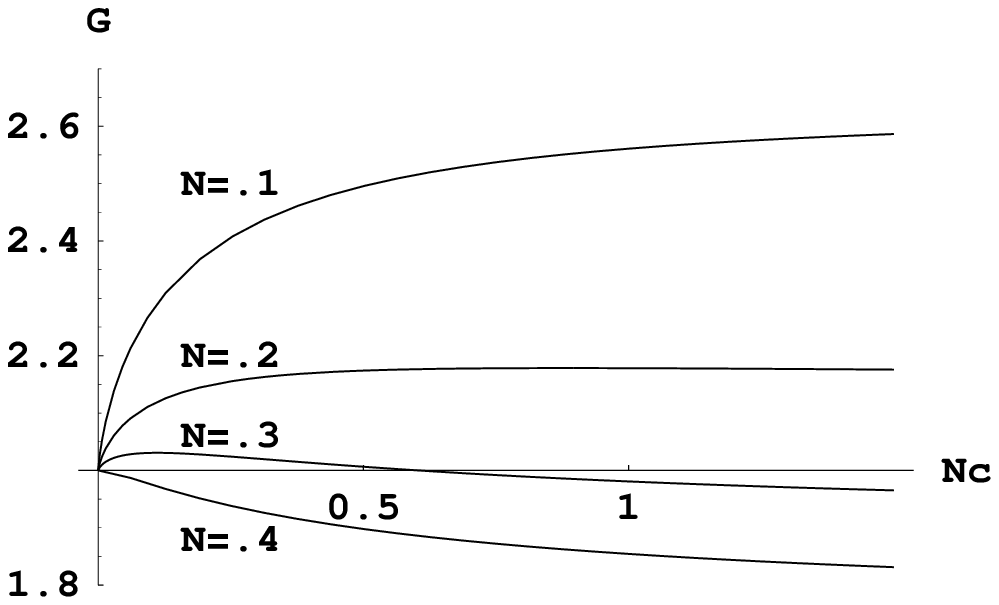} \caption{{\it Gain of
entanglement assistance.}}\small\narrower\noindent Gain
(\ref{gain}) as a function of $N_c$ with $k=1$. Parameter=input
noise $N$.
\end{figure}

\begin{figure}\label{Fig-OutEx}
\epsfxsize=8cm \epsfbox{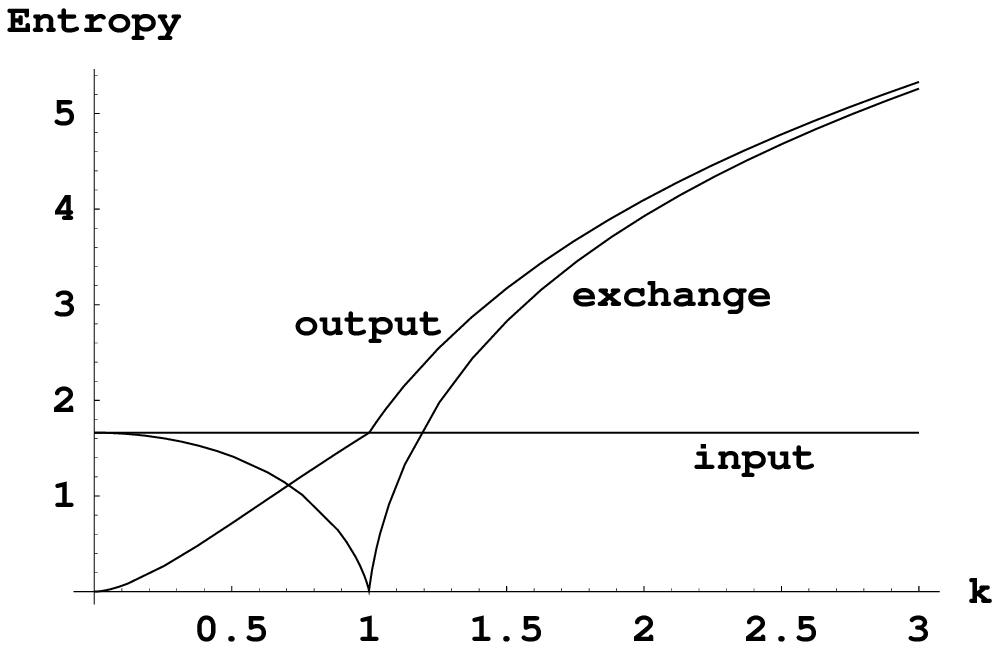} \caption{{\it
Entropies.}}\small\narrower\noindent {\tt output} entropy from
(\ref{out-1mode}), {\tt exchange} entropy from (\ref{exch-1mode})
with $N_c=0$.
\end{figure}

\begin{figure}\label{Fig-capty}
\epsfxsize=8cm \epsfbox{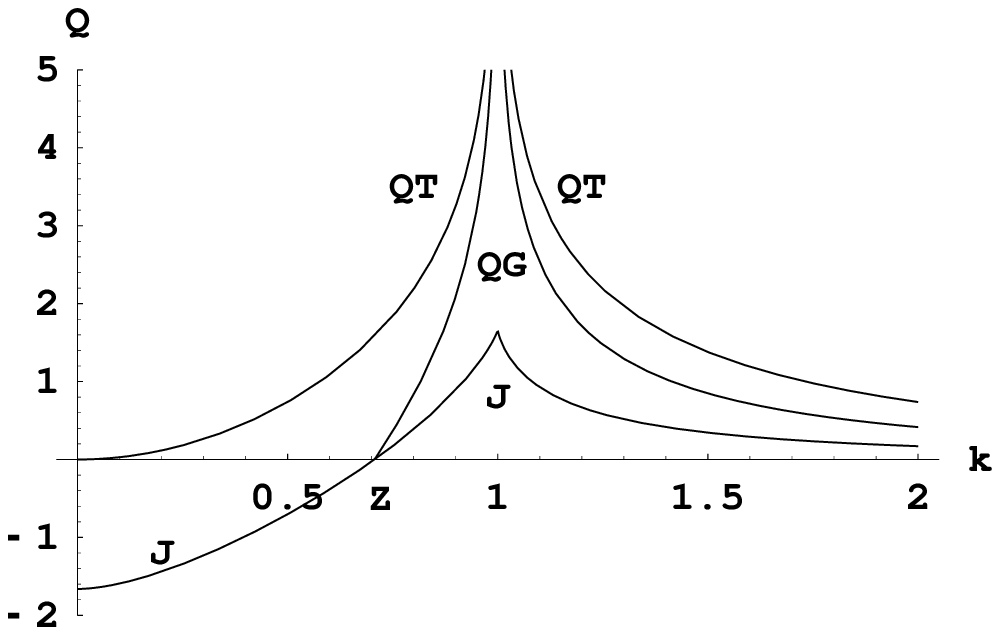} \caption{{\it Bounds for
Quantum Capacity, $N_c=0$.}}\small\narrower\noindent {\tt J}=coherent
information (\ref{coh-1mode}) with $N=.7$;\\ {\tt QG}=$Q_G$= bound maximized
over Gaussians (\ref{QG});\\ {\tt QT}=bound $Q_\Theta$ from
transposition (\ref{1modBound});\\ {\tt Z}= zero at $k=1/\sqrt2$, common
to all curves of type {\tt J}.
\end{figure}

\begin{figure}\label{Fig-capty2} \epsfxsize=8cm
\epsfbox{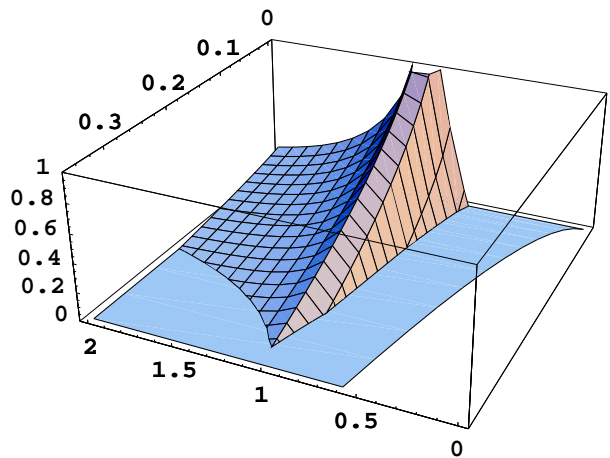} \caption{{\it Gaussian maximized coherent
information} $Q_G(T)$ as function of $k$ and $N_c$. The shaded
area is the area, where $Q_\Theta\geq0$.}. \end{figure}

\end{document}